\documentclass[%
 reprint,
 amsmath,amssymb,
 aps,
]{revtex4-1}

\usepackage{graphicx}
\usepackage{dcolumn}
\usepackage{bm}
\usepackage{color}

\begin{document}

\preprint{APS/123-QED}

\title{A new relation between  the zero of $A_{FB}$ in $B^0 \to K^* \mu^+\mu^-$ and the anomaly in $P_5^\prime$}

\author{Joaquim Matias$^a$}
\author{Nicola Serra$^b$}%
\affiliation{%
$^a$Universitat Autonoma de Barcelona, 08193 Bellaterra, Barcelona \\
$^b$Physik-Institut, Universit\"at Z\"urich, Z\"urich, Switzerland
}%


\date{\today}

\begin{abstract}

We present two exact relations, valid for any dilepton invariant mass region (large and low-recoil) and independent of any effective Hamiltonian computation, between the observables $P_i$ and $P_i^{CP}$ of the angular distribution of the 4-body decay $B \to K^*(\to K\pi) l^+l^-$. These relations emerge out of the symmetries of the angular distribution.  We discuss the implications of these relations under the (testable) hypotheses of no scalar or tensor contributions and no New Physics weak phases in the Wilson coefficients. Under these hypotheses there is a direct relation among the observables $P_{1}$,$P_2$ and $P_{4,5}^\prime$. 
This can be used as an independent consistency test of the measurements of the angular observables. 
Alternatively, these relations can be applied directly in the fit to data, reducing the number of free parameters in the fit. 
This opens up the possibility to perform a full angular fit of the observables with existing datasets. 
An important consequence of the found relations is that a priori two
different measurements, namely the measured position of the zero
($q_0^2$) of the forward-backward asymmetry  $A_{FB}$ 
and the value of $P_5^\prime$ evaluated at this same point, are
related by $P_4^2(q_0^{2})+P_5^2(q_0^{2})=1$. Under the
hypotheses of real Wilson coefficients and $P_4^\prime$ being SM-like, we show that the higher the position of $q_0^{2}$ the smaller
should be the value of $P_5^\prime$ evaluated at the same point. 
A precise determination of the position of the zero of $A_{FB}$
together with a measurement of $P_4^\prime$ (and $P_1$) at  this
position can be used as an independent experimental test of the
anomaly in $P_5^\prime$.  We also point out the existence of upper and lower bounds for $P_1$, namely $P_5^{\prime 2}-1 \leq P_1 \leq 1-P_4^{\prime 2}$, which constraints the physical region of the observables. 


\begin{description}
\item[PACS numbers] 13.25.Hw, 11.30.Er, 11.30.Hv
\end{description}
\end{abstract}

\maketitle


LHCb has performed \cite{LHCbpaper1,LHCbpaper} a measurement of the form factor independent (so called clean) observables \cite{kruger,complete} in the decay $B^0 \to K^*(\to K\pi) \mu^+\mu^-$. These measurements, performed in six independent bins of the dimuon invariant mass squared ($q^2$), were based on a dataset corresponding to an integrated luminosity of 1fb$^{-1}$. Soon after, the first phenomenological analysis of the full set of measurements, at large and low recoil, appeared \cite{thefirst}. 
This analysis had two main conclusions. 
Firstly, it was emphasized that besides a striking 4$\sigma$ deviation in one bin of one observable a set of other less significant deviations (below 3$\sigma$)  were also present in a coherent pattern. Secondly, this pattern pointed to the Wilson coefficient of the semileptonic operator $O_9$ as the main responsible, without excluding possible small contributions from Wilson coefficients of other operators. The connection among those different tensions was shown in Ref.~\cite{thefirst} at the level of operators of an effective Hamiltonian within a specific framework~\cite{qcdf1} to compute QCD corrections. Other analyses that used different approaches \cite{straub,danny, lat} were also presented, including implications for possible NP models~\cite{uli,buras,datta,nazila}.

In the present paper we show that the connection  between the discrepancy in the observables $P_5^{\prime}$ and $P_2$ is deeper and can be proved at a more fundamental level, i.e. using the symmetries of the angular distribution. We point towards a completely new way to test the anomaly in $P_5^\prime$ via a  measurement of the zero in the forward-backward asymmetry {($q^2_0$)} as a key observable. 
At present LHCb measured the zero to be at $q_0^{2}=4.9 \pm 0.9$ GeV$^2$\cite{LHCbpaper1} and our SM prediction is $q_0^{2}=3.95 \pm 0.38$ GeV$^2$. 
The results presented in this paper connect  the values of $P_5^{\prime}$, $P_4^{\prime}$ and $P_1$ when evaluated at $q_0^2$. 

The structure of this paper is the following: in Section~I we recall the symmetry relations between the angular observables and we show how this leads to an exact relation between the clean observables.  In this section, we obtain three results: first, a relation between $P_2$ and the other $P_i$ (and $P_i^{(CP)}$)  observables, second, a new constraint for $P_1$ and third, a relation between the values of different clean observables evaluated at $q_0^2$ (the zero of $P_2$ or $A_{FB}$). In Section II we restrict those relations to the case of no New Physics {(NP)} phases in the Wilson coefficients and all relations simplify considerably.   Then we apply these results to $q^2$ averaged observables. In Section~III we show the implications for future analyses imposing the obtained relations in fits to data.
\section{Exact symmetry relation}
The description of the angular distribution of the decay  $B ^0\to K^*(\to K\pi) \mu^+\mu^-$, if lepton masses and scalar contributions are neglected, 
is completely given by a basis of eight observables \cite{optimizing}
\begin{equation}\label{basis} {\cal O}=\{P_1, P_2, P_3,P_4^\prime,P_5^\prime,P_6^\prime,A_{FB}, d\Gamma/dq^2\}\end{equation}
If lepton masses are considered, two extra observables ($M_{1,2}$  or ${\tilde F_{L,T}}$) have to be added. 
See \cite{complete,swave,optimizing} for definitions. 
In addition the observable $P_8^\prime$ can be used to either substitute one of the $P_i$ observables (for instance $P_3$) or express it in terms of all other observables of the basis. The $P_i$ observables are related to the coefficients ($J_i$) of the angular distribution by
\begin{eqnarray} 
 ({J_{2s}}+ {\bar J_{2s}})&=&\frac{1}{4} N_1, \quad
({ J_{2c}}+{\bar J_{2c}})=-N_2,  \,\,
 \nonumber
\\
 { J_3}+{\bar J_3} &=& \frac{1}{2}  { P_1}   N_1,
\quad
{J_4} + {\bar J_4}= \frac{1}{2}   { P_4'}  N_3, 
\nonumber \\
 { J_5}+{\bar J_5} &=&   { P_5'}  N_3, \quad
 { J_{6s}} + {\bar J_{6s}}=2  { P_2}  N_1,
 \nonumber \\
%
   %
{J_7}+{\bar J_7} &=&-{ P_6'} N_3,
    \quad
 {J_8}+{\bar J_8} =-\frac{1}{2}  { P_8'}  N_3,
    \nonumber \\
   {J_9} +{\bar J_9} &=& -   {P_3}  N_1  \label{defs}     \end{eqnarray}
where $N_{1,2}=\beta^2 F_{T,L} \frac{d \Gamma}{dq^2}$ and $N_3=\beta^2\sqrt{F_T F_L}\frac{d \Gamma}{dq^2}$ with $\beta=\sqrt{1- 4 m_l^2/q^2}$. 
Notice that these expressions are all taken proportional to $\beta^2$ to match the standard definition of the $P_i$ given in \cite{optimizing}.
The set of $P_i^{CP}$, $F_{L,T}^{CP}$ observables are defined using the same Eqs.(\ref{defs})  substituting $J_i + {\bar J_i} \to J_i -{\bar J_i}$ (see \cite{optimizing} for detailed definitions). The coefficients $J_i$  are bilinear functions of the transversity amplitudes $A_{0}^{L,R}, A_{\parallel}^{L,R}, A_{\perp}^{L,R}$ and   the observables $P_i^{(\prime)}$ are ratios of those bilinears. 

Sometime ago, one of us identified four symmetry transformations among the transversity amplitudes that leave the angular distribution invariant \cite{ulrik2}. Working under the {\it hypothesis of no scalar contributions} one can easily solve the transversity amplitudes in terms of the $J_i$ using three of those symmetries (see Sec. 3.3 of \cite{ulrik2}). The remaining fourth symmetry showed up as a relation between the phases  of two of the transversity amplitudes. The following  
 non-trivial consistency relationship between the coefficients of the 
distribution emerges as a byproduct of imposing that the modulus of this relative phase should be one \cite{ulrik2,complete}
\begin{eqnarray}
&&J_{2c}=+ 4\, \frac{\beta_\ell^2 J_{6s} (J_4 J_5 + J_7 J_8) + J_9 (\beta_\ell^2 J_5 J_7 - 4 J_4 J_8)}{16 J_{2s}^{2} -  \left(4 J_3^2+ \beta_\ell^2 J_{6s}^{2} + 4 J_9^2 \right)} \label{relationJ} \\
&&- 2\,\frac{ (2 J_{2s}+ J_3) \left(4 J_4^2+\beta_\ell^2 J_7^2\right) + ( 2 J_{2s} - J_3) \left(\beta_\ell^2 J_5^2+4 J_8^2 \right)}{16 J_{2s}^{2} -  \left(4 J_3^2+ \beta_\ell^2 J_{6s}^{2} + 4 J_9^2 \right)}
 \nonumber
\end{eqnarray}
An identical relation follows for the coefficients ${\bar J}_i$, by simply CP-conjugating Eq.(\ref{relationJ}). 

By using Eq.(\ref{relationJ}) and Eqs.(\ref{defs}) it is possible to write an expression for $P_2$ in terms of the other $P_i^{(\prime)}$ observables.
More precisely, one 
obtains two relations, one between the observables ${\bar P_i}=P_i+P_i^{CP}$ and a second relation between the observables ${\hat P_i}=P_i-P_i^{CP}$. The first relation  is given by
%
%
\begin{widetext}
\begin{eqnarray}  \label{relation}{\bar P_2}=+\frac{1}{2 {\bar k_1}} \bigg[ ( {\bar P_4^\prime} {\bar P_5^\prime} + \delta_{1}) + \frac{1}{\beta }\sqrt{(-1+{\bar P_1} + {\bar P_4^{\prime 2}})(-1-{\bar P_1} + \beta^2 {\bar P_5^{\prime 2}}) +\delta_{2} + \delta_3 {\bar P_1} +\delta_4 {\bar P_1}^2  }\bigg]  \end{eqnarray}
\end{widetext}
where $\delta_i$ are defined in Table \ref{deltas} and where  ${\bar k_1}=1 + {F_L^{CP}}/{F_L}$ and ${\bar k_2}=1+F_T^{CP}/F_T$.
Notice that the existence of this relation is not in contradiction with the fact that the $P_i$ define a basis because Eq.(\ref{relation}) involves 7 of the $P_i$ (and $P_i^{CP}$) but only 6 of them are independent. An identical expression for the ${\hat P_i}$ observables is obtained from Eq.(\ref{relation}) substituting ${\bar P_i} \to {\hat P_i}$ (also inside the $\delta_i$) and ${\bar k_i} \to {\hat k_i}$, where ${\hat k_1}=1 - {F_L^{CP}}/{F_L}$ and ${\hat k_2}=1-F_T^{CP}/F_T$.

\begin{table*}
\caption{\label{deltas}
Definitions of $\delta$ functions in terms of $P_i$ and $P_i^{CP}$ observables.
}
\begin{ruledtabular}
\begin{tabular}{cccc}
$\delta_1= {\bar P_6^\prime} {\bar P_8^\prime} \quad \quad \quad \quad \delta_4=1-{\bar k_1}^2 \quad \quad \quad \quad
\delta_3= (1-{\bar k_1}) {\bar P_4^{\prime 2}} + \beta^2 [(-1+{\bar k_1}) {\bar P_5^{\prime 2}} - {\bar k_1} {\bar P_6^{\prime 2}}]+ {\bar k_1} {\bar P_8^{\prime 2}}$ &
 \\
\hline \\
$\delta_2=-1+{\bar k_1}^2 {\bar k_2}^2 + (1-{\bar k_1} {\bar k_2}) ({\bar P_4^{\prime 2}}+ \beta^2 {\bar P_5^{\prime 2}})  
 - 4 {\bar k_1^2} {\bar P_3^2} + \beta^2 {\bar P_6^{\prime}} {\bar P_8^{\prime}} (2 {\bar P_4^{\prime}} {\bar P_5^{\prime}} + {\bar P_6^{\prime}} {\bar P_8^{\prime}})  
+ {\bar k_1} [\beta^2 {\bar P_6^{\prime}} ( 4 {\bar P_3} {\bar P_5^{\prime}} - {\bar k_2} {\bar P_6^\prime}) - {\bar P_8^{\prime}} (4 {\bar P_3} {\bar P_4^\prime} + {\bar k_2} {\bar P_8^\prime})]
$ & & &
\end{tabular}
\end{ruledtabular}
\end{table*}
Eq.(\ref{relation}) is an exact relation valid for any value of $q^2$.
%
We take "+" sign in front of square root  by consistency with SM,  at low-recoil both solutions ($\pm$) tend to converge at the very endpoint.


From Eq.(\ref{relation}) imposing 
that the argument of the square root is positive, one obtains  the following
restriction on $\bar P_1$ 
\begin{equation} u - \sqrt{u^2+v} \leq {\bar P_1} \leq u + \sqrt{u^2+v} \label{second} \end{equation}
with
\begin{eqnarray} u&=&\frac{1}{2 (1 - \delta_4)} [ (-1+  \beta^2 {\bar P_5^{\prime 2}}) - (-1+{\bar P_4^{\prime 2}}) + \delta_3] \nonumber \\
v&=& \frac{1}{1- \delta_4} [ (-1+{\bar P_4^{\prime 2}})(-1+ \beta^2 {\bar P_5^{\prime 2}})+\delta_2]\end{eqnarray}

Another {\it important consequence}  originates from evaluating Eq.(\ref{relation}) at $q_0^2$. The following relation emerges among the different observables: 
\begin{eqnarray}
[(1 + {\bar P_1}) {\bar P_4^{\prime 2}} + \beta^2 (1- {\bar P_1})  {\bar P_5^{\prime 2}}+ {\bar P_1}^2 + \omega]_{q^2=q_0^2}=1  \quad \label{third}
\end{eqnarray}
where $\omega$ is strongly suppressed and it is defined by
\begin{equation}\omega=\beta^2 \delta_1 (\delta_1 + 2 {\bar P_4^\prime} {\bar P_5^\prime}) - {\bar P_1} (\delta_3 + \delta_4 {\bar P_1}) - \delta_2\end{equation}
%
%
Let us remark that all expressions up to this point are exact, except for the assumptions of no scalar/tensor contributions. In the following we will work within one extra NP hypothesis and one approximation that simplifies considerably the analysis. 

\section{Constrained New Physics and Real Wilson coefficients }

We will assume now that NP does not introduce any new weak phase on Wilson coefficients. This hypothesis implies that $P_i^{CP} \sim 0$ and $F_L^{CP} \sim 0$,  including the small SM contribution. 
Consequently, ${\bar P_i} \to P_i$, ${\hat P_i} \to P_i$ and the two Eqs.(\ref{relation}) become a single equation. This hypothesis can be tested by measuring the $P_i^{CP}$. 
Moreover, taking into account that $J_{7,8,9}$ are functions of ${\rm
  Im} A_i A_j={\rm Im} A_i {\rm Re} A_j +{\rm Im} A_j{\rm Re} A_i $
one can easily see that while $\bar P_{3}$ and $\bar P_{6,8}^\prime$
are ${\cal O}({\rm Im}A_i)$  the $\delta_i$ are further suppressed 
$\delta_i \sim {\cal O}(({\rm Im} A_i)^2, 1-{\bar k_1}, 1-{\hat k_1})$.
\begin{table} 
\begin{ruledtabular}
\begin{tabular}{cccccccc}
&  $\delta_1$ & $\delta_2$ & $\delta_3$ & $\delta_4$ &  \\ \hline \\
\colrule
$|SM|$ & $\lesssim 0.01$ & $\lesssim 0.03$ & $\lesssim 0.01$ & $\lesssim 0.01$ 
 \\
\hline \\
NP & $[-0.03,0.01] $ & $[-0.09,0.01]$ & $[-0.04,0.04]$ & $[-0.03,0.02]$ 
\end{tabular}
\end{ruledtabular} 
\caption{ \label{boundsII} First line corresponds to the $|\delta_i|$ bounds in the SM while second one is the range for $\delta_i$ in presence of NP.}
\end{table}
In the following we  cross check this by using an effective Hamiltonian approach 
 in the SM and in presence of NP. 
 
Even if all the equations discussed up to now are valid for all $q^2$ values, we will focus mainly on the most interesting region $1\leq q^2 \leq 6$ GeV$^2$. In this region the observables ${\bar P_{3}}$, ${\bar P_{6,8}^\prime}$ are approximately bounded in the SM  to be 
$|\bar P_3|\lesssim 5 \times 10^{-3}$, $|{\bar P_6^\prime}| \lesssim 10^{-1}$,  $|\bar P_8^\prime| \lesssim 10^{-1}$.
Given that these observables enter quadratically inside the $\delta_i$, the size of the $\delta_i$ is negligible.
The bounds on the $|\delta_i|$, obtained varying $q^2$ in the 1 to 6 GeV$^2$ region, are given in Table \ref{boundsII}
and the bounds on the relevant combinations entering  Eq.(\ref{relation}) and Eq.(\ref{third}) of previous section  are
 $|\delta_2 + \delta_3 \bar P_1 + \delta_4 {\bar P_1}^2 | \lesssim  0.03$  and $|\omega| \lesssim 0.01$.
The $\omega$ term is evaluated around the $q_0^{2 SM}$  in a 1 GeV$^2$ bin size.
 

 Then, to a very good approximation, Eq.(\ref{relation})  taking  $\delta_i \to 0$ (and ${\bar k}_i \to 1$) simplifies to
\begin{equation} \label{p2} P_2=\frac{1}{2} \left[ P_4^\prime P_5^\prime + \frac{1}{\beta}\sqrt{(-1+P_1 + P_4^{\prime 2})(-1-P_1 + \beta^2 P_5^{\prime 2})} \right] \end{equation}
As Fig.1 (left) shows,  this equation is fulfilled to excellent accuracy in the SM. 

We repeated the analysis of the bounds on $\delta_i$ allowing for the presence of NP in the Wilson coefficients of the dipole and semileptonic operators. We define from now on by NP a range for the Wilson coefficients according to the  (enlarged) pattern found in \cite{thefirst}
\begin{eqnarray}-0.1 &\leq& C_7^{NP} \leq 0.1, \, -2 \leq C_9^{NP} \leq 0, \, -1 \leq C_{10}^{NP} \leq 1 \hfill \nonumber 
\\
-0.1 &\leq& C_7^{\prime} \leq 0.1, \quad -2 \leq C_9^{\prime} \leq 2, \quad -1 \leq C_{10}^{\prime} \leq 1 \quad \,\,\,\, \label{np}\end{eqnarray}
Then, the corresponding range of maximal variation of the  $\delta_i$ terms allowing for NP is given in Table \ref{boundsII}.

The range for the combination of $\delta_i$ terms entering Eq.(\ref{relation}) 
that we obtain in the presence of NP is
 \begin{eqnarray}  &-0.07 \lesssim  \delta_2 + \delta_3 \bar P_1 + \delta_4 {\bar P_1}^2 \lesssim  0.01& \end{eqnarray}
This shows that Eq.(\ref{relation}) is  an excellent approximation, as
Fig.1 (left) illustrates, also in presence of NP.

\begin{table*}
\caption{\label{tab:table3}
Comparison between $P_2$ evaluated in bins and $P_2$ obtained from Eq.(\ref{p2}) assuming its validity in binned form. First row is the difference between the exact result and the result obtained using the relation Eq.(\ref{p2}) in the SM, second and third row are the ranges allowing for New Physics as defined in Eq.(\ref{np}). Last row is the corresponding comparison for one point of NP ($C_9^{\rm NP}=-1.5$). Notice that for small bin size the correction is tiny.
}
\begin{ruledtabular}
\begin{tabular}{cccccccccc}
Point & \textrm{[0.1-2]$^*$}&
\textrm{[2-4.3]}&
\textrm{[4.3-8.68]}
 & \textrm{[1-6]$^*$}
 & \textrm{[1-2]}
  & \textrm{[2-3]}
   & \textrm{[3-4]}
   & \textrm{[4-5]}
 & \textrm{[5-6]}
 \\ \hline \\
$\rm \Delta^{SM}_{exact-relation}$ & $-0.14$ &  $-0.06$  & $-0.03$  & $-0.21$  & $-0.02$ & $-0.02$ & $-0.01$ & $-0.01$ & $-0.01$  \\
\hline \\
${\rm \Delta^{NP}_{exact-relation}}^{upper}$ & $-0.07$ & $-0.02$ & $-0.02$ & $-0.08$  & $+0.00$ & $+0.00$ & $+0.00$ & $+0.00$ & $+0.01$
\\
${\rm \Delta^{NP}_{exact-relation}}^{down}$ &$-0.23$ & $-0.10$ & $-0.09$ & $-0.28$  & $-0.07$ & $-0.04$ &$-0.03$ & $-0.02$ & $-0.04$ \\
\hline \\
${\rm \Delta^{C_9^{NP}=-1.5}_{exact-relation}}$ & $-0.11$ & $-0.04$ & $-0.04$ & $-0.16$  & $-0.01$ & $-0.01$ & $-0.01$ & $-0.01$ & $-0.01$ \\
\end{tabular}
\end{ruledtabular}
\end{table*}
 The  bounds given by Eq.({\ref{second}) also simplify to
 \begin{equation} P_5^{\prime 2} -1 \leq P_1 \leq 1- P_4^{\prime 2} \label{op2} \end{equation}
While this equation is approximate for $\bar P_{1}$ it turns out to be {\it exact} for $P_{1}$,  since it can be also obtained from the simple bound $|P_4|= |P_4^\prime|/\sqrt{1-P_1}\leq 1$ coming from the geometrical interpretation of $P_4$ (see Eq.(16) in \cite{complete}). From $|P_5|=|P_5^\prime|/\sqrt{1+P_1} \leq 1$ one gets the lower bound that is particularly important at low recoil. Also $|P_4^\prime P_5^\prime| \leq 1$ follows. 
%

The evaluation of  $\omega$ in presence of NP around the position of the zero of $A_{FB}$ for each NP point  gives $  -0.01 \lesssim \omega \lesssim 0.07$.
The smallness of this quantity leads to the last important result, namely the condition Eq.(\ref{third}) between the observables evaluated at $q_0^2$ turns out to be

\begin{equation} [P_4^{2}+P_5^{2}]|_{q^2=q_0^2} =1\label{zeroq2x}\end{equation}  assuming $P_1^2(q_0^2)\neq 1$, or in terms of the more interesting $P_{4,5}^\prime$ observables:
\begin{equation} \label{zeroq2first}
[P_4^{\prime 2}+ P_5^{\prime 2}]_{q^2=q_0^2}=1 - \eta(q_0^2) \end{equation}
where 
$$\eta(q_0^2)=[P_1^2 + P_1 (P_4^{\prime 2}-P_5^{\prime 2})]_{q^2=q_0^2} $$
From the geometrical interpretation of the observables $P_{4,5}$ given in Eq.(16-17) of Ref.~\cite{complete} it is evident that they fulfill $|P_{4,5}| \leq 1$. Moreover, in the SM for $q^2 \gg q_0^{2 SM}$ both $|P_{4,5}|$ tend to one, while at $q_0^{2 SM}=3.95$ GeV$^2$ they fulfill Eq.(\ref{zeroq2x}) to an excellent accuracy. In presence of NP, if the zero of $A_{FB}$ appears at a higher position and $P_4$ is SM-like, given that $P_4$ tends to 1 for $q^2>q_0^2$, Eq.(\ref{zeroq2x}) implies that the higher the position of $q^2_0$ the closer to zero the value of $P_5$ at this point should be. The same arguments apply for $P_{4,5}^\prime$ if $q_0^2 > q_0^{2 SM}$ (notice that data prefers a $P_1$ positive in the region of the third bin but the bound coming from $P_4^\prime$ Eq.(\ref{op2}) constrain  $P_1$ to be small in that region), consequently,  $\eta(q_0^2)$ is expected to be small and because of that also the value of $P_5^\prime$ evaluated at $q_0^2$ would tend to a smaller value 
as present data seem to hint.

\subsection{Averaging over \boldmath{$q^2$} }
All previous equations are strictly valid only for a fixed $q^2$
value. However, the measurements are performed as averages in bins of
$q^2$. Since Eq.(\ref{p2}) is not linear in the observables, it is in
general not valid when averaging over $q^2$ regions. 
The only circumstance that would justify the use of this equation when $P_i  \to  \langle P_i \rangle\label{thebinnedform}$, which we will refer to as {\it  binned form of the equation}, would be that all observables were approximately constant  inside the bin. Consequently, this approximation tends to be more valid the smaller the bin size. 
 
 In Table \ref{tab:table3} we evaluate the difference between the
 exact binned result, averaging over $q^2$, and the one obtained using Eq.(\ref{p2}) assuming its validity for $q^2$ average observables, in three cases: a)  SM, b) in presence of NP and c) at the best fit point $C_9^{\rm NP}=-1.5$. In this way we estimate (now using an effective Hamiltonian approach) the maximal shift 
$$\langle P_2 \rangle \to \langle P_2 \rangle + {\Delta^{\rm X}_{\rm exact-relation}}$$
 due to the $q^2$ binning, where $X={\rm SM},{\rm NP}, C_9^{\rm NP}=-1.5$. We compute it using the binning scheme adopted by the experiments~\cite{LHCbpaper1,LHCbpaper,CMSKstmm,Aubert:2008bi,Wei:2009zv,Aaltonen:2011ja} and also for bins of 1 GeV$^2$. 
 \begin{figure}
\includegraphics[width=4.2cm,height=4.5cm]{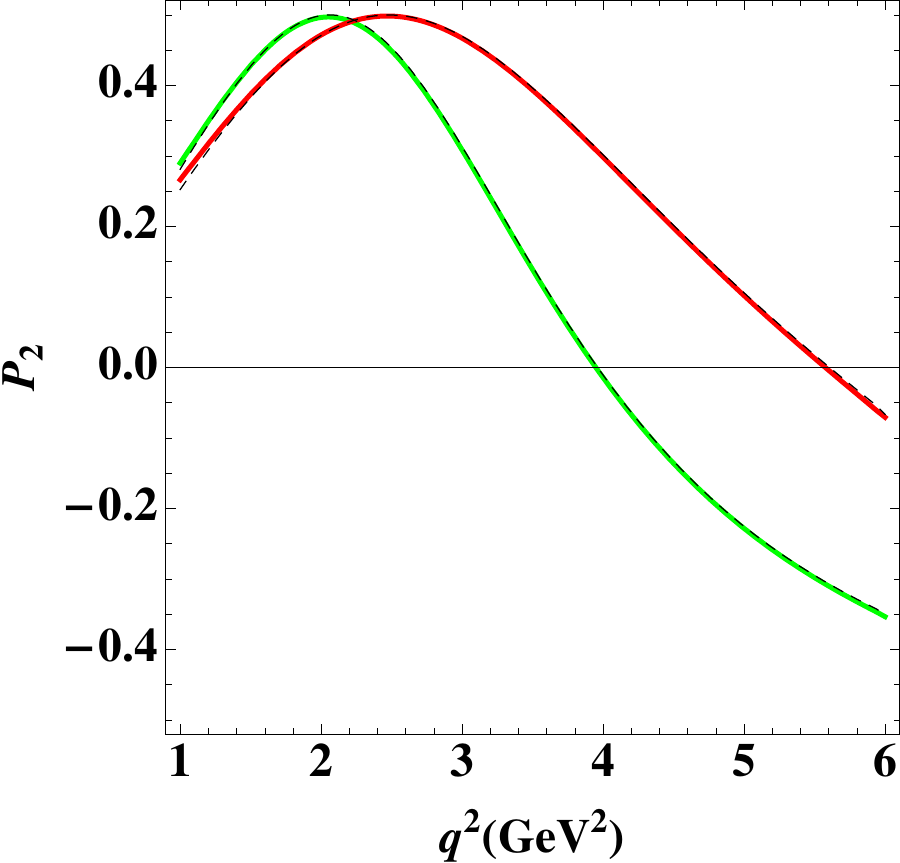}
\includegraphics[width=4.2cm,height=4.5cm]{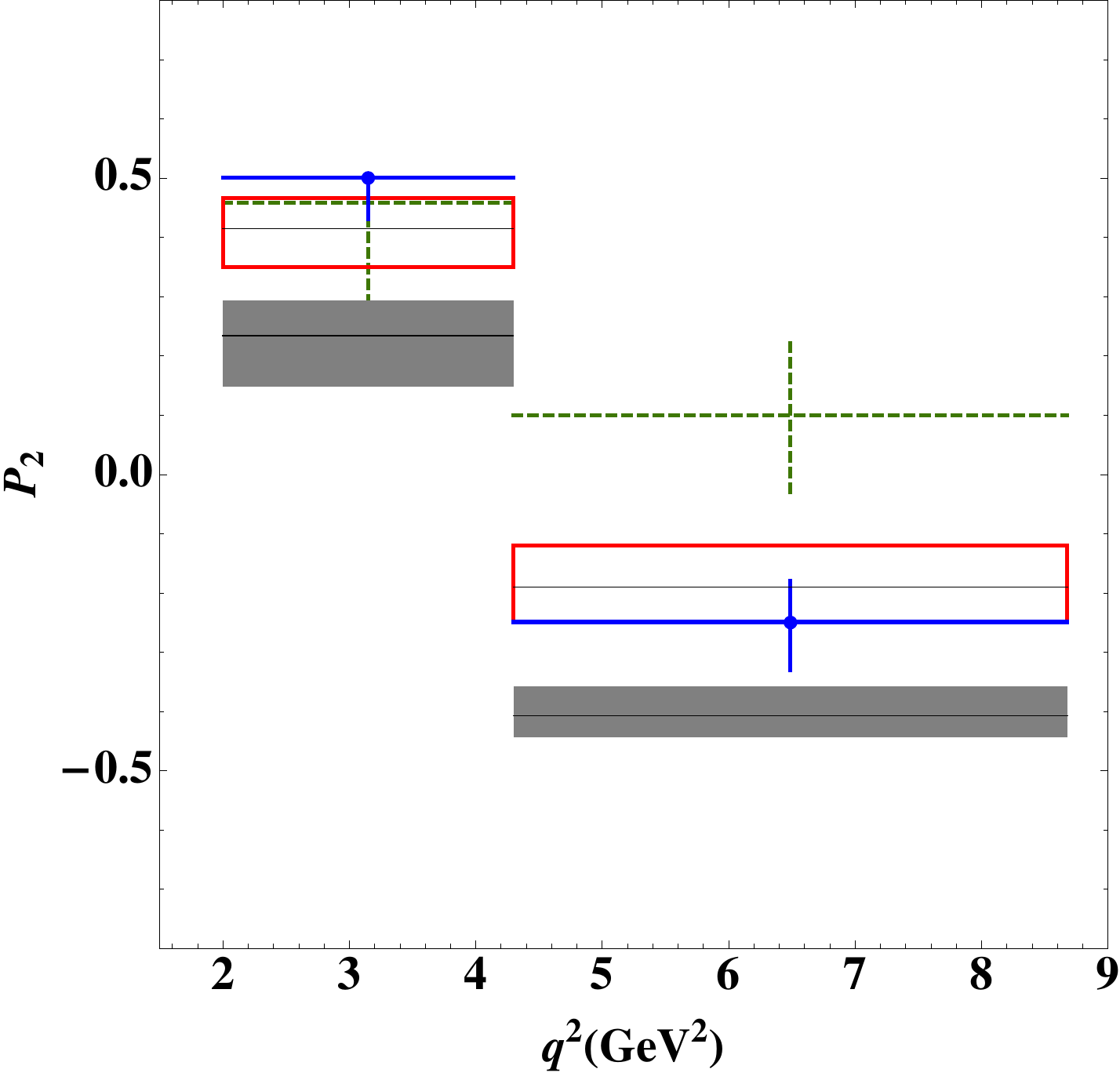}
\caption{(Left) Exact SM curve for $P_2$ (green) and using Eq.(\ref{p2}) (dashed). Exact NP curve for $C_9^{NP}=-1.5$ (red) and using Eq.(\ref{p2}) (dashed). (Right) $P_2$: Gray band is SM, blue cross is the measured value, red box is $C_9^{NP}=-1.5$, green cross dashed is obtained from Eq.(\ref{p2}) using data from $P_1,P_{4,5}^\prime$.}\label{fig:check_comparep2}
\end{figure}
\begin{figure}
\includegraphics[width=8.0cm,height=4.5cm]{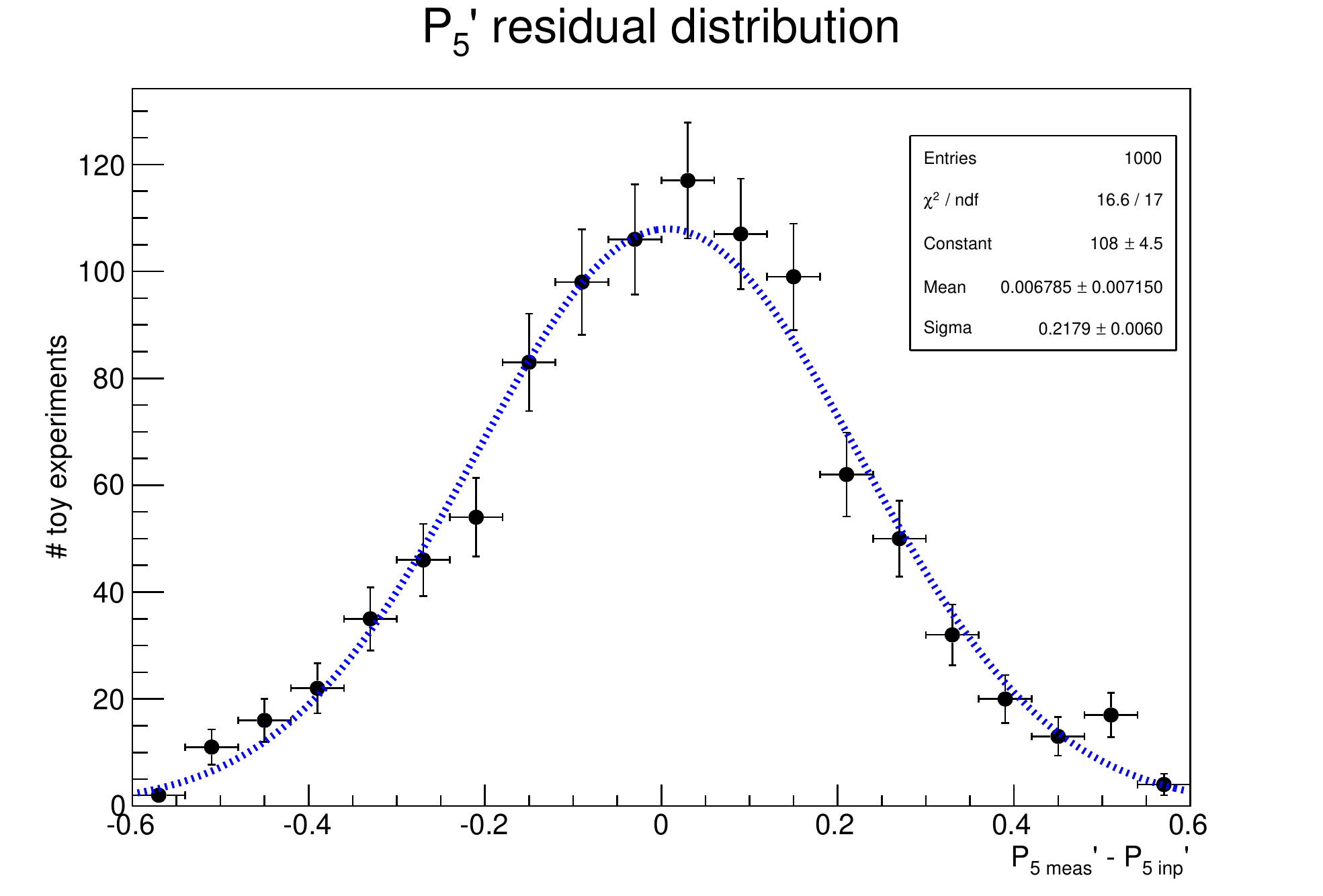}
\caption{Residual distribution of $P_5^{\prime}$ when fitting with 100 events. The fit of a gaussian distribution is superimposed.}\label{residual}
\end{figure}
The conclusions of this analysis are: i) Eq.(\ref{p2}) in its binned form is a good approximation in most of the bins, 
except for the bin [0.1-2]~GeV$^2$ and the bin [1-6]~GeV$^2$, where the shift becomes sizeable ii) the smaller the bin size the smaller the shift, as expected. 
The reason why the shift is so large in the first bin is mainly due to  the $\beta$ factor that varies strongly in this region. 
The bin [1-6]~GeV$^2$ exhibits a large correction given the large size of the bin and the rapid variation of the observables. 
We have also verified that the upper bound for $P_1$ of Eq.(\ref{op2})
is nicely fulfilled in its binned form for all bins given in Table
\ref{tab:table3} and also in presence of NP. Finally,
Eq.(\ref{zeroq2first})  can also be used in binned form. 

\section{Implications for data analysis}

Under the hypotheses of real Wilson coefficients, we performed a frequentist analysis to test the consistency of LHCb measurements. This was done by generating toy experiments taking as input the measured values of $P_1$, $P_4^{\prime}$ and $P_5^{\prime}$~\cite{LHCbpaper}, using Eq.(\ref{p2}) to estimate $P_2$ and comparing the result of this computation with its measured value in the same bin. 
The correction of Table~\ref{tab:table3} for the best fit point ($C_9^{NP}=-1.5$) was applied to correct for the binning effect. 
First of all it should be noted that different $P_i$ observables are
measured independently, and no constraints that the measurements have
to be in the physical region was applied. As a consequence, a fraction
of the generated toy experiments are outside the physical region,
i.e. where the argument of the square root of Eq.~(\ref{p2}) is negative or where Eq.~(\ref{op2}) is not satisfied. 
For the bin [2.0-4.3]~GeV$^2$ about 50\% of the toy experiments fall in the physical region. For the fraction of toy experiments in the physical region, excellent agreement corresponding to 0.2$\sigma$ between the measured value and the value extracted with Eq.~(\ref{p2}) is found. For the bin [1.0-6.0]~GeV$^2$ about 73\% of events fall in the physical region. For these events an agreement corresponding to $0.1\sigma$ is observed. Some tensions are found for the third large recoil bin, with $q^2$ within [4.3-8.68]~GeV$^2$ and the first low recoil bin, with $q^2$ in the region [14.18-16.00]~GeV$^2$. 
In the third large recoil bin only 10\% of events satisfy
Eq.~(\ref{op2}), i.e. the measured value of $P_1$ and $P_4^{\prime}$
are in tension. 
For these events a discrepancy of 2.4$\sigma$ between the value of $P_2$ computed with Eq.~(\ref{p2}) (dashed green cross in Fig.1) and the measured one (solid blue cross in Fig.1) is observed. 
This discrepancy is not surprising since, as was already pointed out in Ref.~\cite{thefirst,conf}, the deviation with respect to the value predicted in the SM in the third bin of $P_5^\prime$ is indeed larger than what the best fit point can explain (notice also that, as discussed in \cite{conf}, other proposed solutions \cite{straub} work significantly worse when evaluated in this bin).
This is reflected in the fact that the value of $P_2$ derived by using
Eq.~(\ref{p2}) for this bin has a discrepancy of 1.9$\sigma$ from the
best fit point, while it has a discrepancy of 3.6$\sigma$ from the SM
prediction (see Fig.~\ref{fig:check_comparep2}). \\
\indent The first low recoil bin has about 70\% of events within the
physical region.  For these events a large discrepancy
  of 3.7$\sigma$ is found between the measured value of $P_2$ and the
  one extracted using "+" sign in Eq.~(\ref{p2}) while agreement is
  found if "-" sign is taken. However one would have expected that both signs would give similar results at low recoil.

\indent Under the assumption of real Wilson coefficients it is possible to use Eq.~(\ref{p2}) directly in the fit, opening up the possibility to have a full fit of the angular distribution with a small dataset. The free parameters in the fit are the observables $F_L$, $P_1$, $P_4^{\prime}$ and $P_5^{\prime}$. The observables $P_{6,8}^{\prime}$ are set to zero, while the observable $P_2$ is determined by using Eq.(\ref{p2}). 
We tested this fit for different values of the observables around the present measured values and we obtained convergence and unbiased pulls with as little as 50 events per bin. This would allow to perform a full fit of the angular distribution with correlations in small bins of $q^2$ with relatively small datasets. Gaussians pulls are obtained with as little as 100 events per bin, as shown in Fig.~\ref{residual} for $P_5^{\prime}$. It is worth remarking that the hypothesis  of no NP weak phases can be tested  by measuring the $P_i^{CP}$ observables.   

In conclusion, the main question we wanted to address with this paper
is if the anomaly in $P_5^{\prime}$ measured by LHCb in the third
large recoil bin is isolated. 
 We found, using only symmetry arguments, that the anomaly should also appear in $P_2$ in a very specific way. 
By means of the newly presented relation  involving $P_{1,2}$, $P_{4,5}^\prime$ 
we have also found that the higher the position of the zero of
$A_{FB}$ the smaller the expected value of  $P_5^\prime$ at this point
(for a SM-like $P_4^\prime$), in agreement with LHCb
measurements. 
These results can be used as an independent consistency test of the measurements of the angular observables. 
A strong constraint on $P_1$ shows that, according to
Eq.(\ref{op2}), experimental values for $P_4^\prime\geq 1$ give no
space for a large positive $P_1$. This rules out those mechanisms
coming from right-handed currents that naturally prefer a large
positive value for $P_1$ in the third large recoil bin, as for
instance [$C_{10}^\prime <0,C_9^\prime >0$] or [$C_{10}^\prime<0, C_{7
  eff}^\prime>0$]. Finally, by using Eq.(\ref{p2}) directly in the
fit to data, under the assumption of no NP weak phases, it is possible to perform a full angular fit with small datasets. 

{\it Acknowledgements:} J.M. acknowledges support from FPA2011-25948, SGR2009-00894. 
N.S. acknowledges the support of the \textit{Swiss National Science
  Foundation}, PP00P2-144674. 
We thank Espen Bowen, Olaf Steinkamp and Patrick
Koppenburg for reading and commenting this document.

\end{document}